\documentclass[draft]{agujournal2019}
\usepackage[inline]{trackchanges}
\usepackage{amsfonts}
\usepackage{amsmath}
\usepackage{amssymb} 
\usepackage{amsthm}  
\usepackage{bm}      
\usepackage{lineno}
\usepackage{lipsum}
\usepackage{soul}
\usepackage{url}
\usepackage{xcolor}
\usepackage{booktabs}
\usepackage{tabularx}
\usepackage{grffile}

\draftfalse
\justifying

\addeditor{Victor}
\addeditor{Gilles}
\addeditor{Marilaure}

\addeditor{grammar}
\addeditor{polishing}

\renewcommand*{\d}[1]{\operatorname{d}\!{#1}}

\journalname{Journal of Advances in Modeling Earth Systems (JAMES)}

\begin{document}
\title{The Physical Limit of Neural Hypoxia Detection in \\ the Black Sea from Satellite Observations}

\authors{
  Victor Mangeleer\affil{1, 2} \hspace{10em}  Luc Vandenbulcke\affil{1} \\
  \vspace{1em}
  Marilaure Grégoire\affil{1} \hspace{2.5em} Gilles Louppe\affil{2}
}

\affiliation{1}{Modelling for Aquatic Systems (MAST), University of Liège}
\vspace{0.75em}
\affiliation{2}{Montefiore Institute of Electrical Engineering and Computer Science, University of Liège}

\correspondingauthor{Victor Mangeleer}{vmangeleer@uliege.be}

\begin{keypoints}

\vspace{0.5em} \item We investigate the prediction of bottom coastal hypoxia from surface satellite-type observations using a deep generative neural network.

\vspace{0.5em} \item We detect 38\% of summer hypoxic events shelf-wide with a precision of 47\% using our neural network emulator.

\vspace{0.5em} \item We emulate the complete three-dimensional sea state with robust accuracy within the mixed layer throughout the year.

\end{keypoints}

\begin{abstract}

\noindent Coastal hypoxia ($O_2 < 63$ [$mmol/m^{3}$]) threatens ocean health worldwide. On continental shelves, summer stratification prevents bottom oxygen consumed by respiration from being renewed, making monitoring essential to protect vulnerable ecosystems and reduce biodiversity loss. Although satellite observations are increasingly available, their potential to infer subsurface oxygen remains largely unexplored. We frame this as a Bayesian inverse problem relating surface observations to the complete three-dimensional physical and biogeochemical states of the Black Sea. Here, we solve it using a deep generative neural network trained on numerical model outputs that provides a tractable and computationally efficient approximation of the true posterior distribution of sea states. We find that accurate state estimation is limited to the mixed layer, because its homogeneity makes surface conditions representative of subsurface states. During summer, we detect 38\% of all hypoxic events shelf-wide with a precision of 47\%. Improving the results will likely require longer assimilation windows or subsurface observations.

\end{abstract}

\section*{Plain Language Summary}

\noindent In coastal waters, nutrients from land-based fertilizers washed in by rivers cause eutrophication, producing organic matter whose decomposition consumes oxygen. When ventilation cannot replenish this loss, oxygen drops below a critical threshold, creating hypoxia that threatens marine ecosystems. Although satellite observations are increasingly available, their potential to directly infer subsurface oxygen levels remains largely unexplored. We address this gap by solving the Bayesian inverse problem that estimates the complete Black Sea states, including oxygen, from satellite surface observations alone. This is challenging since multiple states may produce the same observations. Traditional methods estimate only the most likely state, overlooking rare hypoxic events. Here, we use a deep generative model trained on numerical model outputs to efficiently recover the full range of plausible states, from which oxygen is inferred alongside the other observed variables. We find that surface observations enable accurate state estimation within the mixed layer year-round, because homogeneity within this layer makes surface conditions representative of subsurface states. During summer, we detect 38\% of all hypoxic events shelf-wide, with 47\% of our detections correctly identifying hypoxia. Improving accuracy across depths will likely require longer assimilation windows or subsurface observations (e.g., float measurements).

\newpage \section{Introduction}
\label{sec-main-introduction}
\noindent Hypoxia ($O_2 < 63$ [$mmol/m^{3}$]) is a worldwide phenomenon affecting the bottom of many coastal ocean areas \cite{Schmidtko2017, Breitburg2018, Oschlies2018}. It has detrimental effects on benthic life, particularly crustaceans and bivalves, and more generally on low-mobility organisms that cannot escape \cite{Diaz2008, Vaquer-Sunyer2008}. Hence, accurate and timely estimates of oxygen levels are key to supporting informed management decisions \cite{Capet2013}. The standard approach for estimating oxygen levels and detecting hypoxia relies on numerical ocean models, which simulate the full dynamics of the ocean. In the framework of the Copernicus Marine Environment Monitoring Service (CMEMS) \cite{LeTraon2019}, 10-day forecasts and multidecadal reanalyses are routinely produced for the ocean physics and biogeochemistry, and, in particular, for oxygen. Yet these physics-based models are slow (simulating $\sim$10 days takes hours of walltime), complex to run and not easily amendable by non-experts.

\noindent In operational conditions, satellites continuously observe the ocean surface and measure temperature, salinity, chlorophyll, and sea surface height at ever-increasing temporal, spatial, and spectral resolutions \cite{LeTraon2019}. However, these observations remain confined to the surface and do not directly measure all key ocean indicators such as currents, pH, nutrients, and oxygen levels. No operational system currently exploits them to infer oxygen without first running a numerical simulation and then using the observations to correct it. The real-time availability of these observations thus raises the question of whether surface data alone could enable reliable full state estimation or hypoxia detection without requiring a numerical simulation.

\noindent The possibility of estimating the current sea state, and particularly inferring oxygen levels, from the present-day surface observations is supported by physical and biogeochemical processes \cite{Capet2013}. Bottom hypoxia typically occurs when oxygen consumed by benthic respiration exceeds physical replenishment. This happens under poor ventilation and intense organic matter degradation, most commonly in summer when: ($1$) the water column is stratified, ($2$) bottom temperatures peak, and ($3$) sediments receive organic matter exported after the main surface bloom. This summer decoupling between surface and bottom waters challenges bottom condition retrieval from surface observations. Yet, winter mixing couples surface temperature and chlorophyll to bottom oxygen levels, which condition summer hypoxia. Overall, we argue that the inverse problem linking current surface observations to the subsurface state is physically grounded.

\noindent Diffusion models \cite{Sohl-Dickstein2015, Ho2020, Song2019, Song2020, Karras2022} are well suited to solving inverse problems \cite{Chung2022, Song2023, Rozet2024, Chung2025}. Over the past years, they have been used for the reconstruction of cloud-free data \cite{Barth2024, Yu2025}, downscaling \cite{Wang2024}, surrogate modeling of sea-ice simulations \cite{Finn2024}, and data assimilation \cite{Rozet2023, Andry2025, Martin2025}. Motivated by these successes, we address our inverse problem through a two-stage approach. First, we train a diffusion model on a numerical simulation of the daily evolution of the Black Sea physics and biogeochemistry to learn its dynamics. Then, we use the diffusion model to estimate the current sea state by constraining the generation to satellite-like surface observations. Due to its stochastic nature, the diffusion model enables generation of multiple state estimates that together quantify estimation uncertainty.

\noindent The paper is organized as follows. First, in Section~\ref{sec-main-background}, we introduce diffusion models for solving Bayesian inverse problems. Then, in Section~\ref{sec-main-materials-and-methods}, we present the dataset we learn from, the diffusion model itself, the idealized satellite observation model that constrains it, and our evaluation protocol. In Section~\ref{sec-main-results}, we start by verifying that our diffusion model produces plausible Black Sea states. We then measure how far surface observations constrain the sea state with depth, and what this leaves for hypoxia detection. Finally, in Section~\ref{sec-main-discussion}, we discuss these results, their limitations, and how to go further.

\newpage

\newpage \section{Background}
\label{sec-main-background}
\noindent An inverse problem can be posed as a Bayesian inference problem, where the solution is the posterior distribution $p(x \mid y) \propto p(x) \,  p(y \mid x)$. Here, $p(x)$ is the prior distribution over plausible sea states $x$, and $p(y \mid x)$ is the likelihood describing how observations $y$ relate to a given state $x$. To solve the inverse problem, we start by learning a diffusion model of the prior $p(x)$. Following \citeA{Song2020}, samples $x \sim p(x)$ are progressively perturbed through a diffusion process expressed as a stochastic differential equation (SDE)
\begin{equation}
    \d{x_t} = f_t \, x_t \d{t} + g_t \d{w_t} \, ,
    \label{eq-main-sde-forward}
\end{equation}
where $f_t \in \mathbb{R}$ is the drift coefficient, $g_t \in \mathbb{R}_+$ is the diffusion coefficient, $w_t$ denotes a standard Wiener process, and $x_t$ is the perturbed sample at time $t \in [0, 1]$. Since the SDE is linear in $x_t$, the perturbation kernel from $x$ to $x_t$ is Gaussian and takes the form $p(x_t \mid x) = \mathcal{N}(x_t \mid \alpha_t \, x, \Sigma_t)$, where $\alpha_t$ and $\Sigma_t = \sigma_t^2 \, \mathbb{I}$ are derived analytically from $f_t$ and $g_t$. Remarkably, the forward SDE admits a family of reverse SDEs that also defines a diffusion process \cite{Anderson1982},
\begin{equation}
    \d{x_t} = \left[ f_t \, x_t - \frac{1 + \eta^2}{2} \, g_t^2 \, \nabla_{x_t} \log p(x_t) \right] \d{t} + \eta \, g_t \d{w_t} \, ,
    \label{eq-main-sde-backward}
\end{equation}
where $\eta \geq 0$ controls stochasticity. In practice, the score function $\nabla_{x_t} \log p(x_t)$ is unknown, but can be approximated by a neural network trained via denoising score matching \cite{Vincent2011}. We can draw noise samples from $x_1 \sim p(x_1) \approx \mathcal{N}(x_1 \mid 0, \Sigma_1)$ and gradually denoise them by simulating Eq.~\eqref{eq-main-sde-backward} from $t = 1$ to $0$ to obtain $x_0 \sim p(x_0) \approx p(x)$ using an appropriate discretization scheme \cite{Karras2022}. To sample the posterior $p(x \mid y)$, we need its score, which we can express using Bayes' rule as
\begin{equation}
    \nabla_{x_t} \log p(x_t \mid y) = \nabla_{x_t} \log p(x_t) + \nabla_{x_t} \log p(y \mid x_t) \, ,
    \label{eq-main-posterior-score}
\end{equation}
and plug it into the reverse SDE in Eq.~\eqref{eq-main-sde-backward}. The first term of the right-hand side is efficiently approximated by the neural network, whereas the second term, the perturbed likelihood score $\nabla_{x_t} \log p(y \mid x_t)$, can be approximated under moderate assumptions on the observation process $p(y \mid x)$ \cite{Chung2022, Rozet2024, Daras2024}. Our numerical integration scheme and the choice of hyperparameters are described in Section~\ref{sub-main-diffusion-and-sampling}, while the computation of the prior and likelihood scores is described in Sections~\ref{sub-main-denoising-neural-network} and \ref{sub-main-idealized-observation-model}, respectively. 

\section{Materials and Methods}
\label{sec-main-materials-and-methods}

\subsection{Training Dataset}
\label{sub-main-training-dataset}
\noindent Training the diffusion model requires a large amount of data from which the spatio-temporal dynamics can be learned, but real observations are too few to constitute such a dataset \cite{Skakala2023}. Instead, we use a three-dimensional gridded synthetic dataset from a multidecadal reanalysis, delivered by an ocean modeling system that couples the physical model NEMO \cite{NEMO} and the biogeochemical model BAMHBI \cite{BAMHBI}. The dataset consists of a trajectory $\mathbf{x} = \{x^1, x^2, \dots, x^n\}$, where $n$ is the number of simulated days between 1980 and 2022. Each state $x^i \in \mathbb{R}^{N \times X \times Y \times Z}$ consists of daily-averaged fields, with $N$ the number of variables and $X$, $Y$, and $Z$ the numbers of mesh points along latitude, longitude, and depth. The trajectory has a horizontal resolution of $0.025^{\circ}$ ($\approx$ 2.8 [$km$]) and 59 vertical levels. While it covers the complete Black Sea, we focus our study on the northwestern shelf and only keep the first 32 vertical levels, which gives a total depth of $\approx 145$ [$m$]. This corresponds to a grid of size $128 \times 256 \times 32$. We only keep chlorophyll, salinity, temperature, sea surface height, and oxygen as the emulated variables.

\newpage

\noindent To later simplify the neural network architecture, i.e., by making it possible to use 2D convolutions instead of more expensive 3D ones, we reshape $x^i$ by merging the channel (i.e., variable) dimension $N$ and depth dimension $Z$ into a single axis. Thus, our new state is $x^i \in \mathbb{R}^{C \times X \times Y}$, where $C = (N - 1) \cdot Z + 1$, since sea surface height is a two-dimensional field that occupies a single channel. 

\noindent In addition, we introduce two context variables that provide temporal and spatial information. The temporal context $d \in \mathbb{R}$ represents the day of the year, normalized by $366$ to the interval $[0, 1]$. The spatial context $m \in \mathbb{R}^{3 \times C \times X \times Y}$ encodes the positions and bathymetry of each grid cell using normalized indices in $[0, 1]$. These two variables will be used as additional inputs to the neural network (defined in Section~\ref{sub-main-denoising-neural-network}). 

\noindent We split the dataset into training ($1998$ to $2017$), validation ($2018$ to $2020$), and test ($2021$ to $2022$), discarding data prior to $1998$ because chlorophyll assimilation began around that date, causing a notable shift in chlorophyll dynamics that substantially affected the accuracy and quality of our prior approximation. We preprocess the dataset by ensuring that each state variable has values within its physical range: concentration variables are clipped to zero if negative (in case of a numerical error). Then, we replace all Not-a-Number (NaN) values on the grid representing the land with zeros. We standardize the data as $x = (\hat{x} - \mu_{\text{data}}) / \sigma_{\text{data}}$, where $\hat{x}$ is the original unstandardized data. The mean $\mu_{\text{data}}$ and standard deviation $\sigma_{\text{data}}$ are computed on the training set, for each variable independently at every depth level, because values at different depth levels can differ greatly in magnitude and joint statistics would compress the value range at some levels. We use the same statistics to convert predictions back to physical units. Lastly, we pair each state with its temporal context $d$, so that the training dataset defines the conditional prior distribution $p(x \mid d)$, which the diffusion model learns to approximate.

\subsection{Reverse Diffusion and Sampling}
\label{sub-main-diffusion-and-sampling}
\noindent We adopt the probability flow ordinary differential equation (ODE) formulation of \citeA{Song2020}, which is equivalent to the reverse SDE in Eq.~\eqref{eq-main-sde-backward} when $\eta = 0$. This ODE is defined as
\begin{equation}
    \d{x_t} = \left[ f_t \, x_t - \frac{1}{2} \, g_t^2 \, \nabla_{x_t} \log p(x_t) \right] \d{t} \, .
    \label{eq-main-probability-flow-ode}
\end{equation}
There exists a large variety of possible parameterizations, but we chose to follow the \textit{EDM} parameterization introduced by \citeA{Karras2022}, in which states are perturbed by noise addition only. This corresponds to having $\alpha_t = 1$ and $f_t = 0$, and leaves the diffusion coefficient entirely determined by the noise scheduler, $g_t = \sqrt{2 \, \sigma_t \, \dot{\sigma}_t}$, where $\dot{\sigma}_t$ denotes the time derivative of $\sigma_t$. Therefore, Eq.~\eqref{eq-main-probability-flow-ode} reduces to
\begin{equation}
    \d{x_t} = - \dot{\sigma}_t \, \sigma_t \, \nabla_{x_t} \log p(x_t) \d{t} \, ,
    \label{eq-main-probability-flow-ode-edm}
\end{equation}
leaving the noise scheduler $\sigma_t$ as the only remaining degree of freedom. We chose the custom scheduler $\sigma_t = \sqrt{\sigma_{\min} \cdot \sigma_{\max}} \, \exp\left(\rho \, \operatorname{logit}\left(\varepsilon + (1 - 2\varepsilon) \, t\right)\right)$, where $\sigma_{\min} = 1 \times 10^{-4}$, $\sigma_{\max} = 1 \times 10^{4}$ and $\rho = 2$. The offset $\varepsilon = (\sigma_{\min} / \sigma_{\max})^{1/(2\rho)}$ rescales $t$ so that the schedule spans exactly $\sigma_0 = \sigma_{\min}$ to $\sigma_1 = \sigma_{\max}$. Finally, we chose to solve Eq.~\eqref{eq-main-probability-flow-ode-edm} with a third-order Adams-Bashforth linear multistep integration scheme \cite{Hairer1993, Kidger2022}, which replaces the second-order Heun scheme used in \textit{EDM}. We chose it because, at a slightly higher computational cost, it yields a substantially lower integration error. 

\subsection{Denoising Neural Network}
\label{sub-main-denoising-neural-network}
\noindent We chose the denoiser parameterization $D_\theta$ presented in \textit{EDM}, whose preconditioning terms keep inputs and targets at unit variance across noise levels and limit the amplification of the network errors,
\begin{equation}
    D_\theta(x_t, \sigma_t) = c_{\text{skip}}(\sigma_t) \, x_t + c_{\text{out}}(\sigma_t) \, N_\theta\left(c_{\text{in}}(\sigma_t) \, x_t, \, c_{\text{noise}}(\sigma_t)\right) \, ,
    \label{eq-main-denoiser}
\end{equation}
where the preconditioning coefficients $c_{\text{skip}}(\sigma_t)$, $c_{\text{out}}(\sigma_t)$, $c_{\text{in}}(\sigma_t)$ and $c_{\text{noise}}(\sigma_t)$ are taken from the \textit{EDM} column of the first table of \citeA{Karras2022} and where $N_\theta\left(x_t, \, \sigma_t \right)$ is the neural network (a U-Net, described below), with trainable parameters $\theta$. Before passing through the denoiser, the context variables $d$ and $m$ are projected and concatenated to the noised state $x_t$. For notational convenience, we continue referring to this extended representation as $x_t$. The denoiser is trained to estimate the posterior mean $\mathbb{E}[x \mid x_t]$ by minimizing
\begin{equation}
    \mathbb{E}_{x \sim p(x \mid d)} \, \, \mathbb{E}_{t \sim \mathcal{U}(0,1)} \, \, \mathbb{E}_{\epsilon \sim \mathcal{N}(0, \sigma_t^2 \mathbb{I})} \, \, \lambda(\sigma_t) \, \lVert D_\theta(x_t, \sigma_t) - x \rVert^2_2,
    \label{eq-main-loss}
\end{equation}
where $x_t = x + \epsilon$ and $\lambda(\sigma_t) = 1 + 1 / \sigma_t^2$. We improve this objective with two modifications. First, the error is masked to train on sea regions only. Second, the errors at each depth level are inversely weighted by the proportion of sea surface area at that depth relative to the total Black Sea surface. This ensures balanced learning across depths. Once trained, the score of the perturbed prior follows from the denoiser through Tweedie's first-order formula \cite{Robbins1956, Efron2011},
\begin{equation}
    \nabla_{x_t} \log p(x_t \mid d) = \frac{\mathbb{E}[x \mid x_t] - x_t}{\sigma_t^2} \, \approx \frac{D_\theta(x_t, \sigma_t) - x_t}{\sigma_t^2} \, ,
    \label{eq-main-tweedie}
\end{equation}
which we can plug back into Eq.~\eqref{eq-main-probability-flow-ode-edm} to sample from our prior approximation $p_{\theta}(x \mid d)$. As the network $N_\theta$, we use a U-Net \cite{UNetOriginal} with noise level (instead of time) conditioning \cite{Ho2020}. The encoder compresses the input state $x_t$ into latent representations across four stages ($512$, $1024$, $1024$, $2048$ channels), while the decoder reconstructs the output at the original resolution. Each stage contains two or three sequential residual blocks, each combining a skip connection with input normalization, noise level $\sigma_t$ modulation, and two convolutions. The final model was trained for $8$ consecutive days using $32$ A100 ($40$ [$GB$]) GPUs, with the training hyperparameters listed in Table~\ref{tab-app-training-hyperparameters}. The complete architecture hyperparameters are reported in Table~\ref{tab-app-neural-network-architecture}.

\subsection{Idealized Observation Model and Likelihood Score}
\label{sub-main-idealized-observation-model}
\begin{figure}[!hb]
    \noindent\includegraphics[width=\textwidth]{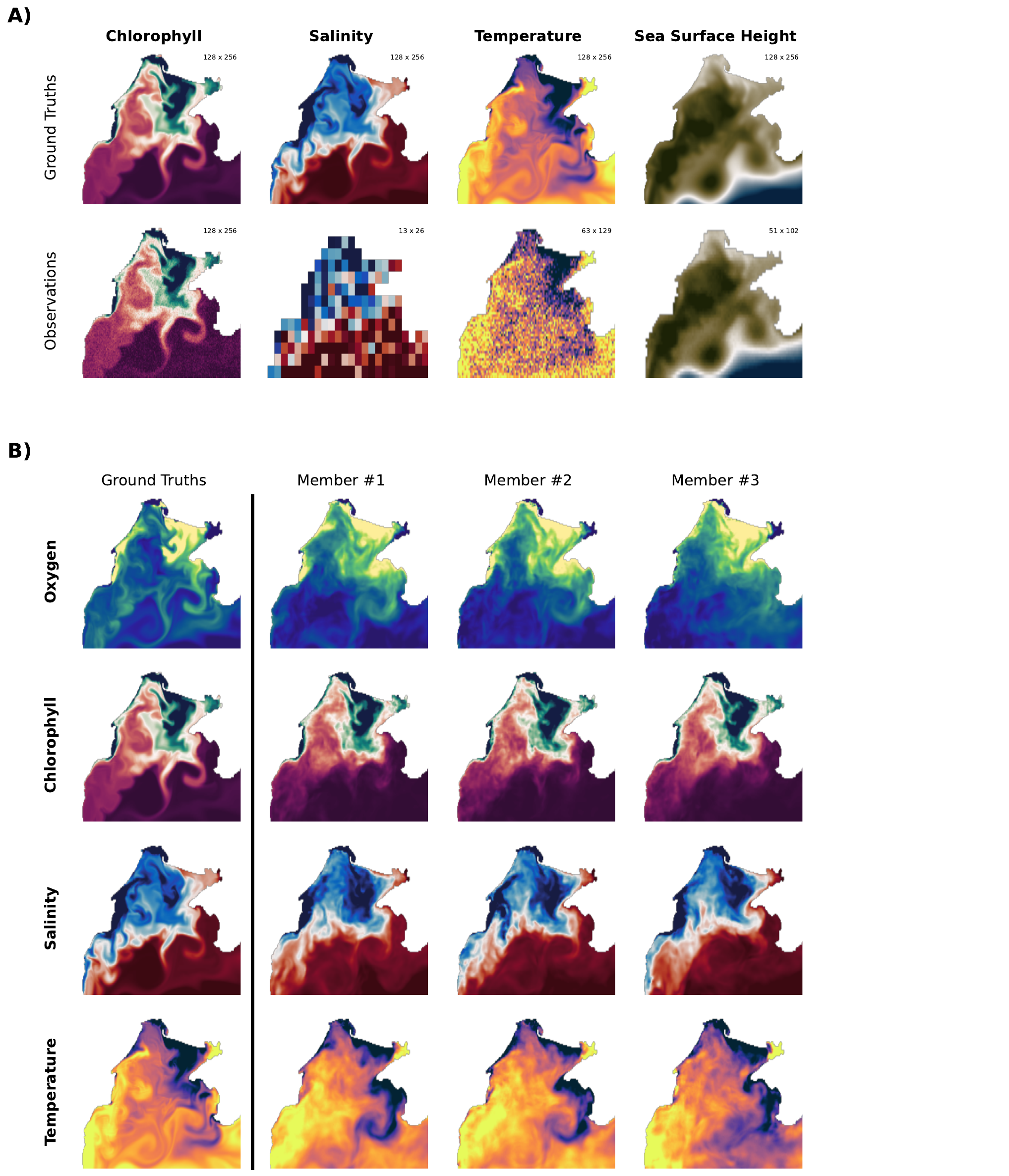}
    \caption{(\textbf{A}) Example of Black Sea surface observations $y$ generated using the idealized observation model in Eq.~\eqref{eq-main-observation-process-approximation} for an arbitrary state $x$ from the test dataset. All state variables are observed except oxygen. (\textbf{B}) Posterior ensemble members $x \sim p_{\theta}(x \mid y, d)$ for a subset of emulated variables at depth $8.2$ [$m$]. Additional ensemble members are shown in Figs.~\ref{fig-app-posterior-oxygen}--\ref{fig-app-posterior-temperature}.}
    \label{fig-main-observations}
\end{figure}

\noindent In order to compute the perturbed likelihood score $\nabla_{x_t} \log p(y \mid x_t)$ in Eq.~\eqref{eq-main-posterior-score}, we first need to define our observation process $p(y \mid x)$. We approximate it as a linear Gaussian forward model capturing the essential characteristics of satellite measurements
\begin{equation}
    p(y \mid x) \approx \mathcal{N}(y \mid A \,x + \mu_{S}, \, \Sigma_{S}),
    \label{eq-main-observation-process-approximation}
\end{equation}
where $A$ maps from simulation space onto the observation space of each variable, using bilinear interpolation when the real satellite product is coarser and the identity when it is finer, and $\mu_{S}$ and $\Sigma_{S} = \sigma_{S}^2 \, \mathbb{I}$ are the bias and covariance matrix modeling the satellite observational error. The Gaussian form is standard in data assimilation \cite{Evensen2009,Carrassi2017} and physically motivated by the aggregation of independent instrument errors \cite{Desroziers2005}. We further assume a diagonal $\Sigma_{S}$, as the full error covariance is rarely known in practice and hard to estimate \cite{Bouttier1999}. Both $\mu_{S}$ and the diagonal entries of $\Sigma_{S}$ are calibrated from the product documentation of the Black Sea Level-3 satellite products distributed by the Copernicus Marine Service, and are reported in Table~\ref{tab-app-observation-parameters} together with the observation resolutions. Finally, we do not use spatial masks to partially hide the surface, due to the difficulty of generating patterns that match actual satellite coverage \cite{Dolinar2024}. Examples of observations $y$ generated by our idealized observation model in Eq.~\eqref{eq-main-observation-process-approximation} are shown in Fig.~\ref{fig-main-observations}A.

\noindent We compute $\nabla_{x_t} \log p(y \mid x_t)$ with the
\textit{Moment Matching Posterior Sampling} (MMPS) algorithm of \citeA{Rozet2024}. They approximate $p(x \mid x_t)$ by a Gaussian distribution whose first two moments are obtained using our denoiser $D_\theta$. More specifically, we have
\begin{equation}
    \mathbb{E}[x \mid x_t] \approx D_\theta(x_t, \sigma_t)
    \qquad \text{and} \qquad
    \mathbb{V}[x \mid x_t] \approx \sigma_t^2 \, \nabla_{x_t} D_\theta(x_t, \sigma_t) \, .
    \label{eq-main-mmps-moments}
\end{equation}
Since our observation model is linear and Gaussian, the perturbed likelihood is Gaussian as well and its score is available in closed form,
\begin{equation}
    \nabla_{x_t} \log p(y \mid x_t) \approx \nabla_{x_t} \mathbb{E}[x \mid x_t]^{\top} A^{\top}
    \left( \Sigma_{S} + A \, \mathbb{V}[x \mid x_t] \, A^{\top} \right)^{-1}
    \left( y - \mu_{S} - A \, \mathbb{E}[x \mid x_t] \right).
    \label{eq-main-mmps}
\end{equation}

\noindent In practice, we solve the linear system in Eq.~\eqref{eq-main-mmps} using a single GMRES iteration \cite{Saad1986} that requires only vector-Jacobian products with $D_\theta$. Following Eq.~\eqref{eq-main-posterior-score}, adding the prior score of Eq.~\eqref{eq-main-tweedie} and the likelihood score of Eq.~\eqref{eq-main-mmps} gives the posterior score, which we plug into Eq.~\eqref{eq-main-probability-flow-ode-edm} to sample from $p_{\theta}(x \mid y, d)$. In Fig.~\ref{fig-main-observations}B, we show a few posterior ensemble members $x \sim p_{\theta}(x \mid y, d)$ for a subset of the emulated variables at a fixed depth generated solely from the observations in Fig.~\ref{fig-main-observations}A.

\vfill  \newpage

\subsection{Evaluation Protocol}
\label{sub-main-evaluation-protocol}
\noindent Experiments fall into two categories: (1) we validate our diffusion prior and (2) we analyze posterior samples. For (1), we draw samples $x \sim p_{\theta}(x \mid d)$ and assess how well they reproduce the state dynamics of the training dataset. For (2), we first generate idealized observations $y$ from each state $x$ in the test dataset using our idealized observation model. Afterwards, we use these observations to draw samples $x \sim p_{\theta}(x \mid y, d)$ and create an ensemble of $128$ state estimates $x$. Each member is generated independently by solving Eq.~\eqref{eq-main-probability-flow-ode-edm} discretized in $256$ timesteps. Since all $128$ members are independent, they can be generated simultaneously, one per A100 ($40$ [$GB$]) GPU, so the full ensemble is produced in $1.25$ [$min$] total.

\subsection{Metrics and Experiments}
\label{sub-main-metrics-and-experiments}

\noindent First, the \textbf{1-Wasserstein distance} \cite{Peyr2018} quantifies the dissimilarity between two one-dimensional probability distributions. It is a non-negative scalar, and the smaller the value, the more similar the two distributions (see Eq.~\eqref{eq-app-wasserstein}). We compute three distances per variable, denoted $W_1$, $W_2$ and $W_3$. To begin with, $W_1$ measures the irreducible sampling variability between two finite approximations of $p(x \mid d)$ and serves as our target. $W_2$ measures how well our learned prior $p_{\theta}(x \mid d)$ approximates the reference prior $p(x \mid d)$. Lastly, $W_3$ measures the distance between $p(x \mid d)$ and the marginal distribution $p(x)$. A well-approximated prior should satisfy $W_1 \approx W_2 < W_3$, meaning its error falls within the irreducible sampling variability. Second, the \textbf{spread-skill ratio} (SSR) \cite{Fortin2014} combines the \textbf{skill} (see Eq.~\eqref{eq-app-skill}), i.e. the root mean square error of the ensemble mean, and the \textbf{spread} (see Eq.~\eqref{eq-app-spread}), i.e. the square root of the ensemble variance, into a single metric defined as
\begin{equation}
    \text{SSR} = \sqrt{\frac{M+1}{M}} \, \, \frac{\text{Spread}}{\text{Skill}},
    \label{eq-main-ssr}
\end{equation}
where $M$ is the ensemble size. A well-calibrated ensemble has an SSR close to $1$, meaning its uncertainty appropriately reflects its error. An SSR below $1$ indicates over-confidence (too narrow relative to error), while an SSR above $1$ indicates under-confidence (too dispersed relative to error). Third, we evaluate our hypoxia detection performance using three standard classification metrics (see Eqs.~\eqref{eq-app-balanced-accuracy}--\eqref{eq-app-recall}). The \textbf{precision} measures the fraction of predicted hypoxic voxels that are truly hypoxic, the \textbf{recall} measures the fraction of true hypoxic events correctly detected, and the \textbf{balanced accuracy} measures the overall skill at distinguishing hypoxic from non-hypoxic regions. All three range from $0$ to $1$, with higher values indicating better results. Fourth, for a grid column located at $(a,b)$, the \textbf{hypoxia density} measures the percentage of daily states over a given period for which hypoxia was detected at any depth level,
\begin{equation}
    \text{Hypoxia Density}_{a,b} = \frac{100}{N_T} \sum_{i=1}^{N_T} 
    \mathbf{1}\!\left[\min_{z} \, x_{a,b,z}^i < \tau \right],
    \label{eq-main-hypoxia-density}
\end{equation}
where $N_T$ is the number of days in the period, $i$ indexes daily states, $z$ indexes depth levels, $\tau$ is the hypoxia threshold, and $\mathbf{1}[\cdot]$ is the indicator function. Therefore, this metric quantifies how frequently hypoxia occurs at a given location over a period of time, without indicating where it is found in the water column.

\section{Results}
\label{sec-main-results}

\subsection{Diffusion Prior Validation}
\label{sub-main-prior-validation}
\begin{figure}[!ht]
    \noindent\includegraphics[width=\textwidth]{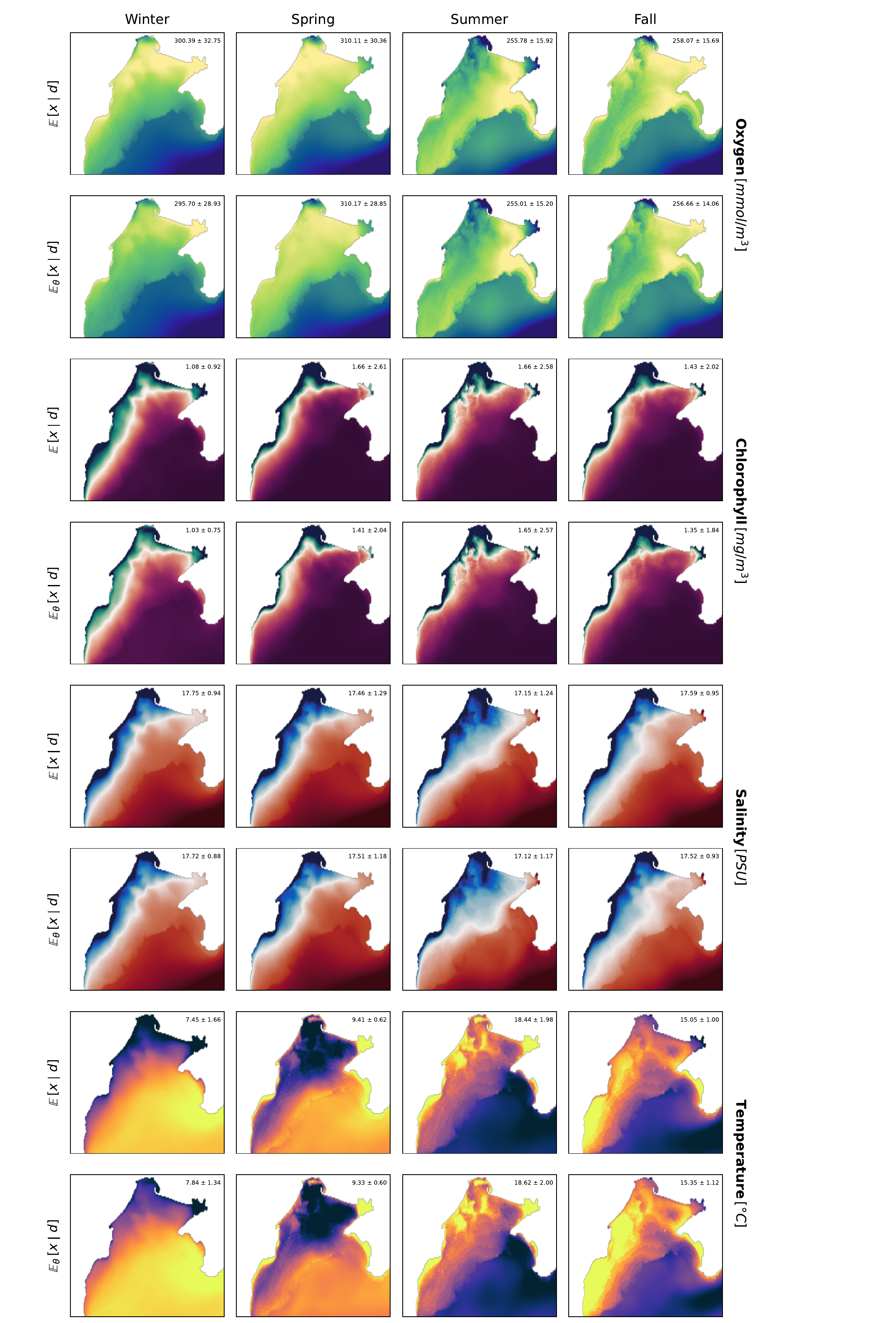}
    \caption{Seasonally and vertically averaged emulated variables (sea surface height is not shown) for reference $\mathbb{E}[x \mid d]$ and approximation $\mathbb{E}_{\theta}[x \mid d]$, with spatial mean and standard deviation reported in each panel. Close agreement confirms accurate learning of seasonal dynamics.}
    \label{fig-main-prior-validation-seasonal-mean}
\end{figure}

\noindent We start by analyzing the emulated average seasonal dynamics shown in Fig.~\ref{fig-main-prior-validation-seasonal-mean}. The prior captures the summer extension of the Danube's plume, which spreads across the entire shelf as river discharge peaks. In summer, low-salinity waters occupy the shelf and productive coastal waters extend offshore. The temperature seasonal cycle, lowest in winter-spring and highest in summer-fall, is particularly well represented in the Danube plume.

\newpage \vfill  \newpage

\noindent Oxygen dynamics are linked to temperature through solubility effects and to river discharge through photosynthesis and respiration, especially at depth. The emulated oxygen field matches the reference in terms of spatial patterns and intensity. In all fields, the influence of river discharges from the Danube, Dniester and Dnieper is well represented, and sub-mesoscale structures are present at the frontal interface. 

\begin{figure}[!h]
    \noindent\includegraphics[width=0.99\textwidth]{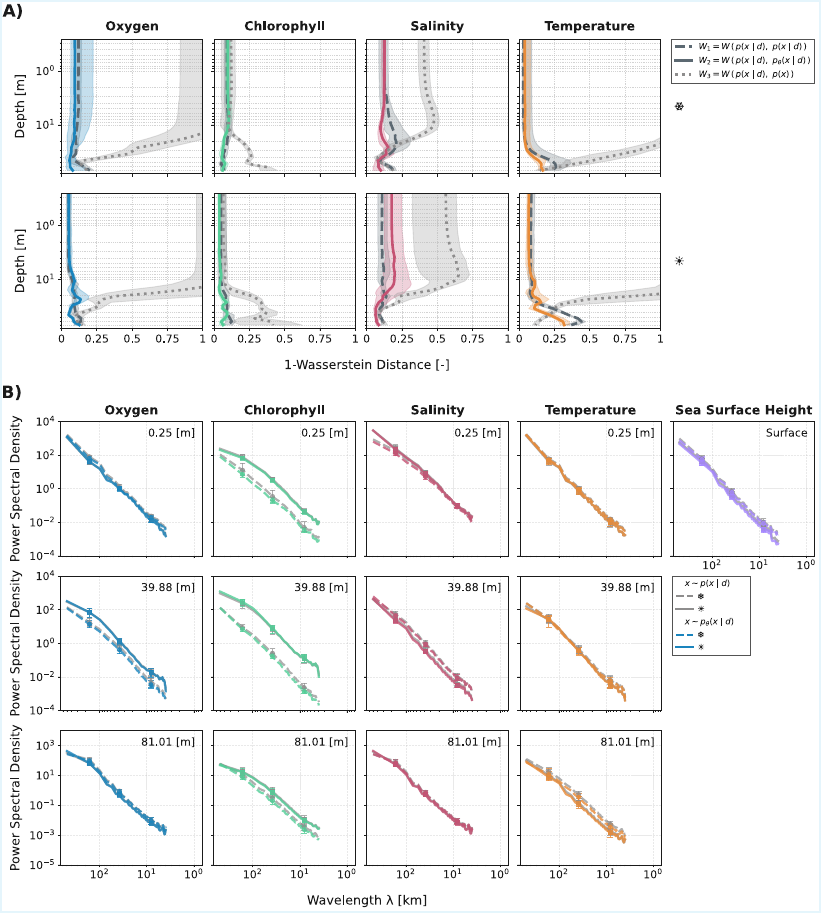}
    \caption{(\textbf{A}) 1-Wasserstein distances $W_1$ (grey dashed), $W_2$ (colored solid), and $W_3$ (grey dotted), defined in Section~\ref{sub-main-metrics-and-experiments}, showing median and quantiles (25\% and 75\%) per depth level for each state variable. The closer $W_2$ is to $W_1$, the better the prior approximation. (\textbf{B}) Power spectral density showing median and quantiles (25\% and 75\%) for reference samples $x \sim p(x \mid d)$ (grey) and generated samples $x \sim p_{\theta}(x \mid d)$ (colored), for winter (dashed) and summer (solid). The closer the colored curves are to the grey curves, the better the prior reproduces spatial structures.}
    \label{fig-main-prior-analysis}
\end{figure}

\noindent Then, in Fig.~\ref{fig-main-prior-analysis}A, $W_2$ stays small and close to $W_1$ across all variables, depths and seasons, meaning our approximation error falls within the irreducible sampling variability. The gap between $W_1$ and $W_3$ also follows the seasonal structure of the water column. 

\noindent Within the mixed layer, the well-mixed upper part of the water column, oxygen, salinity and temperature vary strongly between daily and yearly timescales, whereas chlorophyll does not, because episodic blooms create similar variability at both. Below the mixed layer the pattern reverses, chlorophyll diverging through seasonal subsurface formation and destruction while oxygen and salinity converge in the stable anoxic and halocline regions. Finally, in Fig.~\ref{fig-main-prior-analysis}B, the power spectral densities align at all wavelengths in both winter and summer, confirming that spatial structures are reproduced across scales. Additional visualizations of prior samples are shown in Fig.~\ref{fig-app-prior-samples-0.25}.

\subsection{Impact of Surface Observations on State Estimation}
\label{sub-main-state-estimation}
\begin{figure}[!ht]
    \noindent\includegraphics[width=0.98\textwidth]{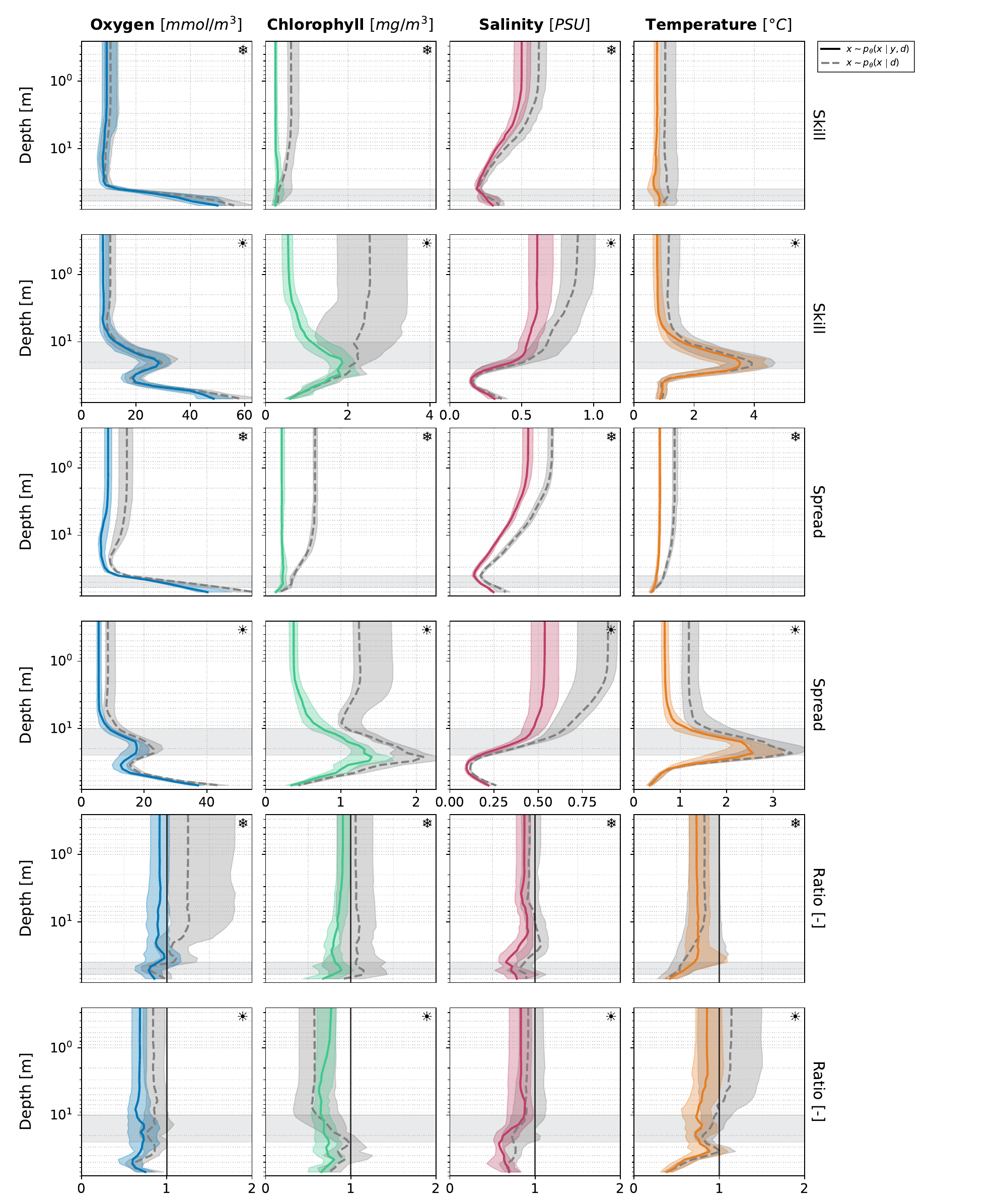}
    \caption{Skill (estimation error), ensemble spread, and spread-skill ratio median and quantiles (25\% and 75\%) for ensembles generated with (colored) and without (grey) surface observations $y$, for winter and summer. Surface observations reduce error and spread for all variables, including unobserved oxygen, with the largest improvement for oxygen in summer. The horizontal grey band marks the depth range where the mixed layer ends.}
    \label{fig-main-ssr}
\end{figure}

\noindent We assess the impact of surface observations $y$ on state estimation by comparing ensembles of estimates $x \sim p_{\theta}(x \mid y, d)$ against $x \sim p_{\theta}(x \mid d)$. We use the spread-skill ratio to quantify whether constraining to surface observations improves estimation accuracy across all depths, only at certain depths, or provides no additional benefit. Results are in Fig.~\ref{fig-main-ssr}. 

\noindent First, we observe that surface observations reduce estimation error ($1$st and $2$nd rows) for all variables in both winter and summer, whether observed or unobserved (oxygen). For observed variables, the reduction is significant, whereas for oxygen, the improvement is modest overall but most pronounced in summer. For all emulated variables, this improvement only holds within the mixed layer. Once its maximum extension depth is reached (somewhere in the grey horizontal region), both error curves start to align. This means that below this depth, our state estimates are as good as if we did not use any surface information. This depth limitation is consistent with the physical expectation that surface conditions directly constrain mainly the well-mixed upper layer. For the spread ($3$rd and $4$th rows), surface observations reduce ensemble spread, reflecting increased certainty about the state. Finally, for the spread-skill ratios ($5$th and $6$th rows), they remain between $0.5$ and $1$ throughout the whole depth column, indicating that all ensembles are over-confident. Many factors may contribute to this over-confidence. First, the diagonal $\Sigma_{S}$ approximation (Eq.~\eqref{eq-main-observation-process-approximation}) treats spatially correlated observations as independent, which over-counts the information they carry. Second, we sample with the deterministic probability flow ODE, whose only source of randomness is the initial noise. Third, MMPS approximates $p(x \mid x_t)$ by a Gaussian and truncates the linear system to a single iteration, which pulls samples towards the posterior mean.

\subsection{Inferring Oxygen Dynamics and Hypoxia}
\label{sub-main-hypoxia}
\noindent Hypoxia predominantly occurs near the shelf bottom \cite{Capet2013}, which is well below the mixed layer. We established that surface observations provide no predictive skill at those depths. Indeed, the estimates reduce to a daily climatology. However, we quantify our ability to detect hypoxia along the whole shelf water column, as some hypoxic events also occur within the mixed layer. To do so, we use classification metrics as a function of depth under two detection thresholds ($63$ and $80$ [$mmol/m^{3}$]), and we also examine a spatial hypoxia density map alongside precision maps. For both analyses, we compare idealized observations $y$ against high-resolution observations $y^*$ at full domain resolution (examples of $y^*$ are shown in Fig.~\ref{fig-app-hr-observations}). The results are shown in Fig.~\ref{fig-main-hypoxia}.

\subsubsection{Satellite-Resolution Observations}
\label{sub-sub-main-hypoxia-idealized}
\noindent Using idealized observations and the standard threshold, the metrics averaged over the mixed layer depth ($0$ to $25$ [$m$]) are $65.0$\% accuracy, $49.7$\% precision, and $34.0$\% recall. By adjusting the detection threshold to $80$ [$mmol/m^{3}$], we increase the number of regions classified as hypoxic but also increase false alarms. Despite this trade-off, the overall detection performance improves: accuracy increases by $1.5$\%, recall by $9.8$\%, while precision decreases by only $4.2$\%. Averaged over the water column ($0$ to $80$ [$m$]) with the adjusted threshold, we reach $67.7$\% accuracy, $46.9$\% precision, and $38.1$\% recall. 

\vfill  

\begin{figure}[!ht]
    \noindent\includegraphics[width=0.95\textwidth]{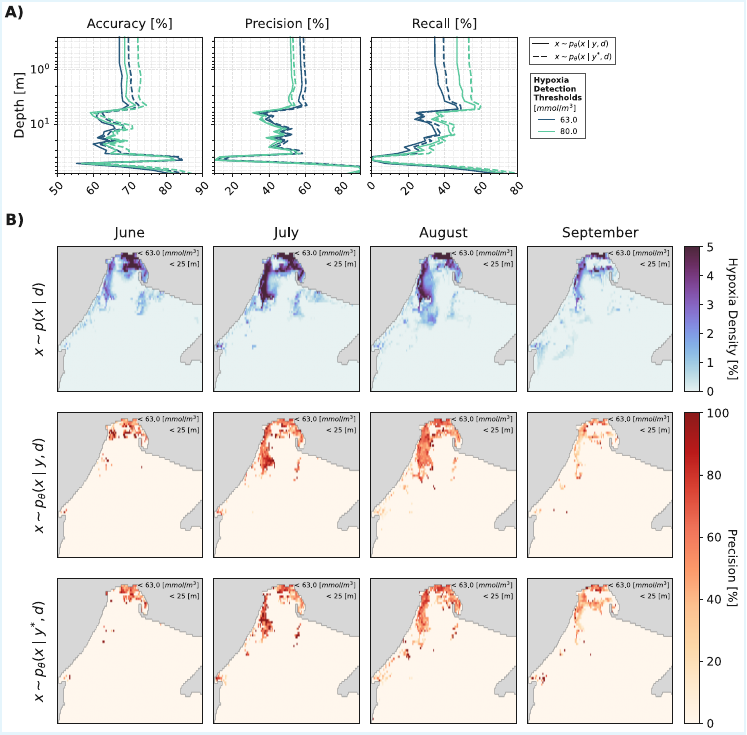}
    \caption{(\textbf{A}) Mean classification performance metrics (accuracy, precision, recall) as a function of depth, averaged spatially and temporally over June to September for both test years. Results are shown for two detection thresholds and two observation types: idealized satellite observations $y$ (solid lines) and high-resolution observations $y^*$ (dashed lines). (\textbf{B}) Spatial analysis over the first $25$ [$m$] for each summer month. The first row shows the ground truth hypoxia density, indicating where hypoxia is consistently present. The second and third rows show the monthly average precision per voxel, indicating where the model correctly predicts hypoxia along the water column.}
    \label{fig-main-hypoxia}
\end{figure}

\noindent Moving to our hypoxia spatial analysis, the ground truth hypoxia density (Fig.~\ref{fig-main-hypoxia}B, first row) reveals a persistent high-density region along the northwestern shelf that expands throughout summer, alongside sparsely distributed low-density regions ($\le 1\%$). The monthly average precision per voxel (Fig.~\ref{fig-main-hypoxia}B, second row) shows that correct predictions are spatially concentrated in the high-density regions of the ground truth, while sparse, low-density regions go undetected. Thus, surface observations provide enough information to identify regions where hypoxia is likely, but not enough to capture rare, spatially isolated events.

\newpage  

\subsubsection{High-Resolution Observations}
\label{sub-sub-main-hypoxia-high-resolution}
\noindent We find that high-resolution observations $y^*$ improve all classification metrics globally, which demonstrates that observation quality directly affects the vertical oxygen distribution and that a margin of improvement exists. However, the spatial precision maps (Fig.~\ref{fig-main-hypoxia}B, third row) closely resemble those from idealized observations (second row). This means that better surface observations refine our oxygen estimates where hypoxia was already detected, but detections remain concentrated in high-density regions. Sparse, spatially isolated hypoxic events are located far from these regions and go undetected regardless of observation quality. Therefore, we argue that improving surface observations alone cannot resolve the detection gap for rare hypoxic events: additional observations are needed, either a longer assimilation window of past surface observations or in-situ measurements.

\section{Discussion and Conclusions}
\label{sec-main-discussion}
\noindent We trained a deep generative neural network on a reanalysis dataset to solve the Bayesian inverse problem relating satellite surface observations of temperature, salinity, chlorophyll, and sea surface height to the three-dimensional physical-biogeochemical states of the northwestern Black Sea. This includes oxygen, which is inferred solely from these surface-observed variables without any direct oxygen observation. This approach captures the full posterior distribution, estimates uncertainty, enables fast inference, and can leverage any combination of observations at various resolutions without retraining the neural network.

\noindent We start by validating our diffusion prior, confirming that $p_{\theta}(x \mid d)$ accurately approximates $p(x \mid d)$ by analyzing seasonal means. We then conduct two sets of experiments using idealized satellite observations. The first reveals that below the mixed layer, surface observations provide no predictive skill, a direct consequence of the physical decoupling between the surface and the sea interior. The second provides a quantitative assessment of hypoxia detection over the whole shelf water column. This physical limit is reflected in the detection metrics: averaged over the whole shelf water column during summer, we achieve only $38.1$\% recall and $46.9$\% precision. The spatial precision maps analysis further shows that higher-resolution surface observations improve detection but cannot overcome this information barrier, suggesting the need for longer assimilation windows or in-situ measurements.

\noindent We acknowledge that the test period (2021 to 2022) lies only four years beyond the end of training (2017), during which Black Sea oxygen dynamics may not have changed substantially. Evaluating on more temporally distant data would better assess whether detection performance remains consistent as oxygen conditions evolve. Potential improvements include refining the observation model to account for observation correlations, extending the framework to the full Black Sea, and integrating additional observation sources such as Argo and BGC-Argo floats to better capture the spatial diversity of hypoxia.

\section*{Open Research Section}
\noindent The source code, a subset of the preprocessed data, and the diffusion model weights are available on Zenodo at \url{https://zenodo.org/records/19628325}, \url{https://zenodo.org/records/19629825}, and \url{https://zenodo.org/records/19632543}. The official project repository is available at \url{https://github.com/VikVador/Poseidon}. The complete physical and biogeochemical datasets used in this study are distributed via the Copernicus Marine Service and respectively available at \url{https://doi.org/10.48670/mds-00356} \cite{CMEMS_PHY}, and \url{https://doi.org/10.48670/mds-00372} \cite{CMEMS_BGC}. \nocite{PoseidonCode, PoseidonWeights, PoseidonData}

\section*{Conflict of Interest declaration}
\noindent The authors declare there are no conflicts of interest for this manuscript.

\acknowledgments
M.~Grégoire and G.~Louppe co-supervised this work and contributed equally. We thank all members of MAST and SAIL for their helpful discussions and support. We benefited from the computational resources provided by the Consortium des Équipements de Calcul Intensif en Fédération Wallonie-Bruxelles (CECI), funded by the Fonds de la Recherche Scientifique (F.R.S.-FNRS) under grant 2.5020.11. All co-authors acknowledge support from the ESA project MITHO (MESA -- MultIple THreats on Ocean health -- MiTHo Project, contract no.\ 4000142100/23/I-DT), the Horizon Europe BioGeoSea project (under grant agreement no.\ 101216427), and the Copernicus Marine Service (CMEMS, BS-MFC). We acknowledge the use of Claude Sonnet 4.6 (Anthropic) to improve the manuscript's clarity.

\newpage \bibliography{references}

\newpage \appendix

\section{Numerical Configuration}
\label{app-configuration}

\vfill \begin{table}[!ht]
    \caption{Hyperparameters for training the denoiser $D_\theta$.}
    \centering
    \label{tab-app-training-hyperparameters}
    \renewcommand{\arraystretch}{1.5}
    \begin{tabularx}{\columnwidth}{@{} l c @{}}
        \specialrule{1pt}{0pt}{0pt}
        Optimizer                      & SOAP \cite{Vyas2024}   \\
        Learning rate                  & $2 \times 10^{-4}$   \\
        Weight decay                   & 0.0 \\
        Scheduler                      & constant \\
        Gradient norm clipping         & 1.0 \\
        Batch size per GPU             & 4 \\
        Gradient accumulation steps    & 4 \\
        Total optimization steps       & $256\,000$ \\
        \specialrule{1pt}{0pt}{0pt}
    \end{tabularx}
\end{table}

\vfill \begin{table}[!ht]
    \caption{Hyperparameters of the neural network $N_\theta$.}
    \centering
    \label{tab-app-neural-network-architecture}
    \renewcommand{\arraystretch}{1.5}
    \begin{tabularx}{\columnwidth}{@{} l c @{}}
        \specialrule{1pt}{0pt}{0pt}
        Residual blocks per level      & [2, 2, 3, 3]   \\
        Channels per level             & [512, 1024, 1024, 2048]   \\
        Kernel size                    & $3 \times 3$ \\
        Attention                      & None \\
        Modulation features            & 64 \\
        Activation                     & SiLU   \\
        Normalization                  & LayerNorm   \\
        Dropout                        & 0.0 \\
        \specialrule{1pt}{0pt}{0pt}
        \textbf{Total parameters}      & $7.5 \times 10^8$   \\
        \specialrule{1pt}{0pt}{0pt}
    \end{tabularx}
\end{table}

\vfill \begin{table}[!ht]
    \caption{Bias $\mu_{S}$, standard deviation $\sigma_{S}$ and observation resolutions of the idealized observation model (Eq.~\eqref{eq-main-observation-process-approximation}). The native resolution is that of the Level-3 satellite product, while the effective resolution is the observation grid onto which $A$ maps the simulation state.}
    \centering
    \label{tab-app-observation-parameters}
    \renewcommand{\arraystretch}{1.5}
    \begin{tabularx}{\columnwidth}{@{} l c c c c @{}}
        \specialrule{1pt}{0pt}{0pt}
        & $\mu_{S}$ & $\sigma_{S}$ & Native resolution & Effective resolution \\
        \hline
        Chlorophyll [$mg/m^3$]           & 0.17  & 0.40 & $256 \times 326$ & $128 \times 256$ \\
        Salinity [$PSU$]                 & $-0.07$ & 0.91 & $13 \times 26$   & $13 \times 26$   \\
        Temperature [$^{\circ}C$]        & 0.09  & 0.51 & $63 \times 129$  & $63 \times 129$  \\
        Sea Surface Height [$10^{-3}m$]  & 0.00  & 1.22 & $51 \times 102$  & $51 \times 102$  \\
        \specialrule{1pt}{0pt}{0pt}
    \end{tabularx}
\end{table}

\newpage \section{Formulas}
\label{app-formulas}
\noindent Here, we mathematically define the remaining metrics used in Section~\ref{sub-main-metrics-and-experiments}. All of them are computed on a spatial grid of size $X \times Y$, independently at each depth level. First, the \textbf{1-Wasserstein distance} between two distributions $P$ and $Q$ is estimated from spatial samples of size $n = X \times Y$ as
\begin{equation}
    W(P, Q) = \frac{1}{n} \sum_{i=1}^{n} \left| p_{(i)} - q_{(i)} \right| \, ,
    \label{eq-app-wasserstein}
\end{equation}
where $p_{(i)}$ and $q_{(i)}$ denote the $i$-th smallest values of the samples drawn from $P$ and $Q$, both sorted in ascending order. Second, we consider $K$ ensembles of $M$ members each, and denote by $x^{k}_{a,b}$ the ground truth of the $k$-th ensemble at grid point $(a,b)$ and by $\hat{x}^{k,m}_{a,b}$ its $m$-th member. The \textbf{skill} measures how far the ensemble mean lies from the ground truth, through their root mean square error,
\begin{equation}
    \text{Skill} = \sqrt{\frac{1}{K X Y} \sum_{k=1}^{K} \sum_{a=1}^{X} \sum_{b=1}^{Y} \left( x^{k}_{a,b} - \frac{1}{M} \sum_{m=1}^{M} \hat{x}^{k,m}_{a,b} \right)^{2}} \, ,
    \label{eq-app-skill}
\end{equation}
while the \textbf{spread} measures how far the members lie from their own mean, through the square root of the ensemble variance,
\begin{equation}
    \text{Spread} = \sqrt{\frac{1}{K X Y} \sum_{k=1}^{K} \sum_{a=1}^{X} \sum_{b=1}^{Y} \frac{1}{M-1} \sum_{m=1}^{M} \left( \hat{x}^{k,m}_{a,b} - \frac{1}{M} \sum_{n=1}^{M} \hat{x}^{k,n}_{a,b} \right)^{2}} \, .
    \label{eq-app-spread}
\end{equation}
Lastly, the \textbf{balanced accuracy}, the \textbf{precision} and the \textbf{recall} are computed from the counts of true positives (TP), true negatives (TN), false positives (FP) and false negatives (FN),
\begin{equation}
    \text{Balanced Accuracy} = \frac{1}{2}\left(\frac{\text{TP}}{\text{TP}+\text{FN}} +
    \frac{\text{TN}}{\text{TN}+\text{FP}}\right),
    \label{eq-app-balanced-accuracy}
\end{equation}
\begin{equation}
    \text{Precision} = \frac{\text{TP}}{\text{TP}+\text{FP}},
    \label{eq-app-precision}
\end{equation}
\begin{equation}
    \text{Recall} = \frac{\text{TP}}{\text{TP}+\text{FN}}.
    \label{eq-app-recall}
\end{equation}

\newpage \section{Prior Samples}
\label{app-samples-prior}

\begin{figure}[!h]
    \centering
    \includegraphics[width=0.99\textwidth]{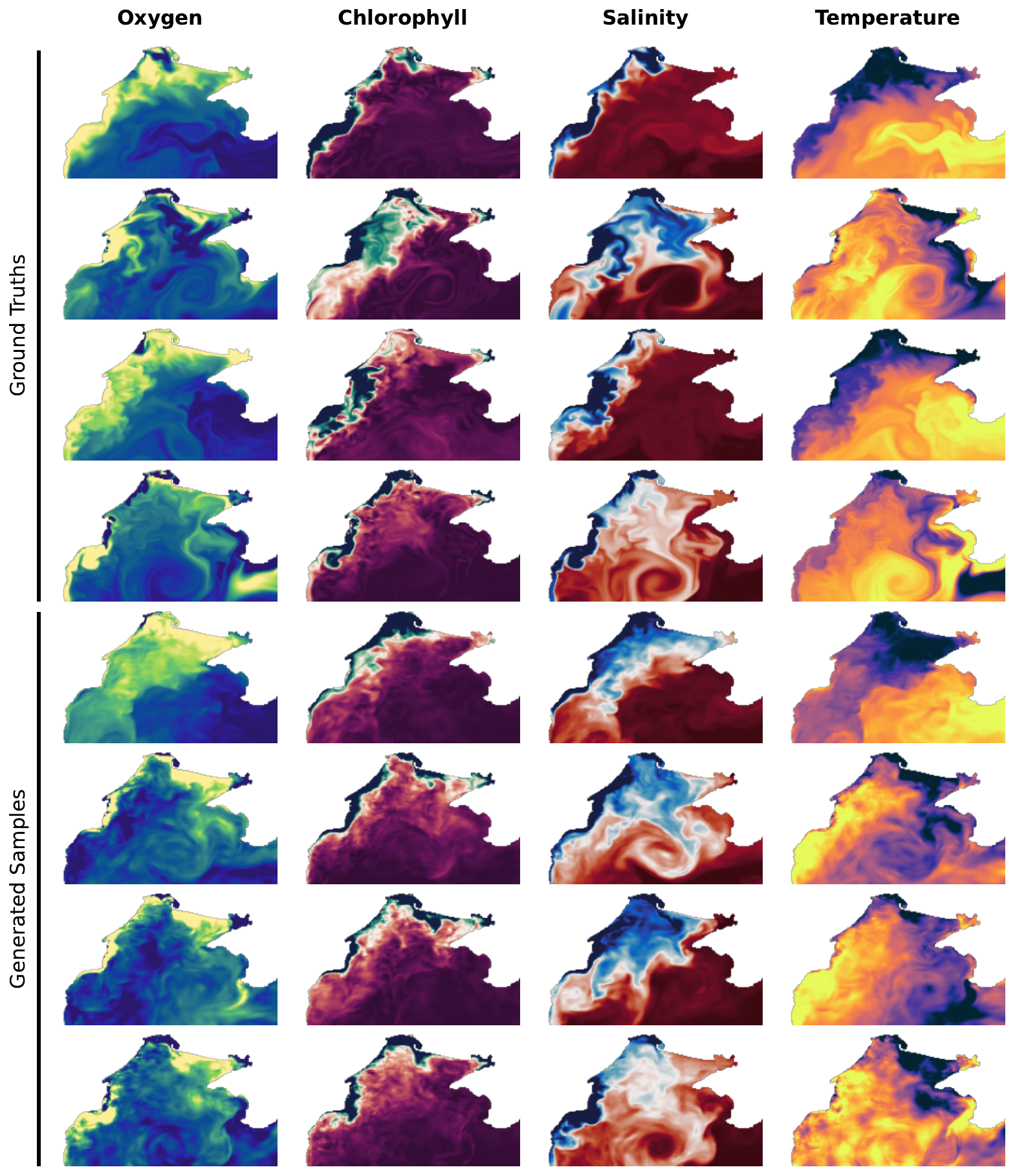}
    \caption{Black Sea states at depth $0.25$ [$m$] for various days $d$ and all state variables except sea surface height. The first four rows show reference states $x \sim p(x \mid d)$. The last four rows show generated samples $x \sim p_{\theta}(x \mid d)$.}
    \label{fig-app-prior-samples-0.25}
\end{figure}

\newpage
\section{Posterior Samples}
\label{app-samples-posterior}

\begin{figure}[!h]
    \noindent\includegraphics[width=0.99\textwidth]{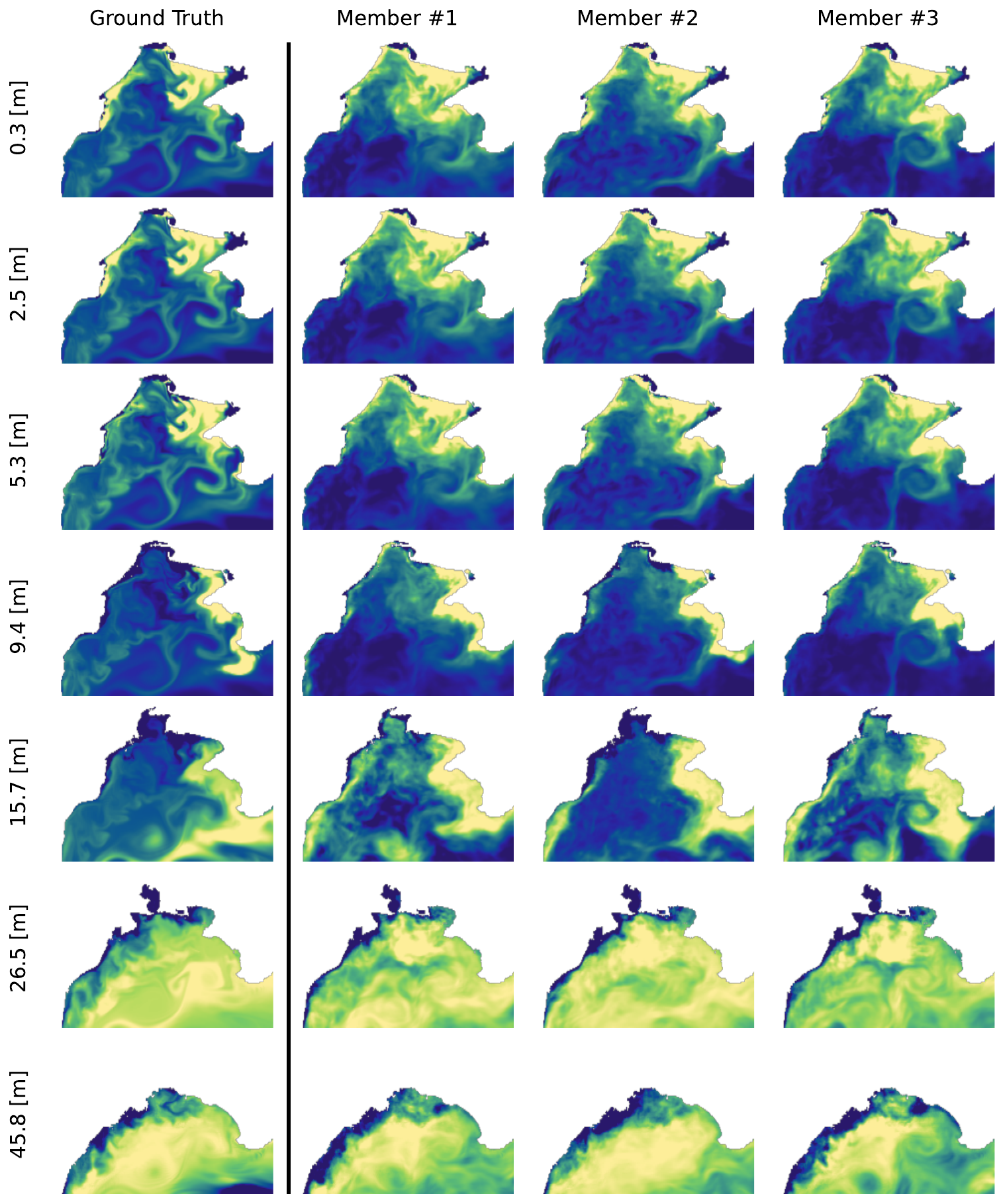}
    \caption{Posterior ensemble members $x \sim p_{\theta}(x \mid y, d)$ for \textbf{oxygen} at various depths, using idealized observations $y$ shown in Fig.~\ref{fig-main-observations}A. The first column shows the ground truth state. The remaining columns show different ensemble members.}
    \label{fig-app-posterior-oxygen}
\end{figure}

\newpage \begin{figure}[!h]
    \noindent\includegraphics[width=0.99\textwidth]{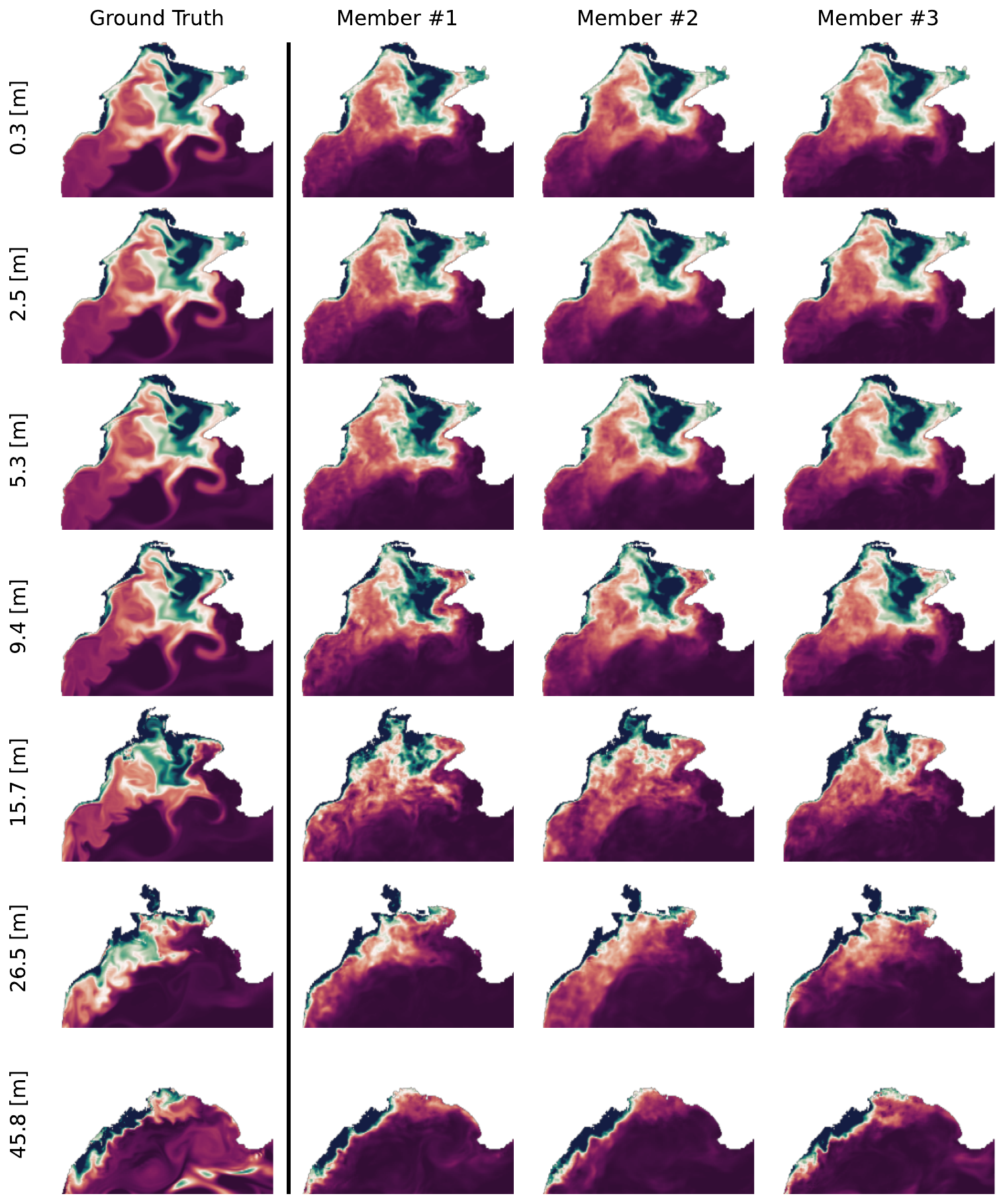}
    \caption{Posterior ensemble members $x \sim p_{\theta}(x \mid y, d)$ for \textbf{chlorophyll} at various depths, using idealized observations $y$ shown in Fig.~\ref{fig-main-observations}A. The first column shows the ground truth state. The remaining columns show different ensemble members.}
    \label{fig-app-posterior-chlorophyll}
\end{figure}

\newpage \begin{figure}[!h]
    \noindent\includegraphics[width=0.99\textwidth]{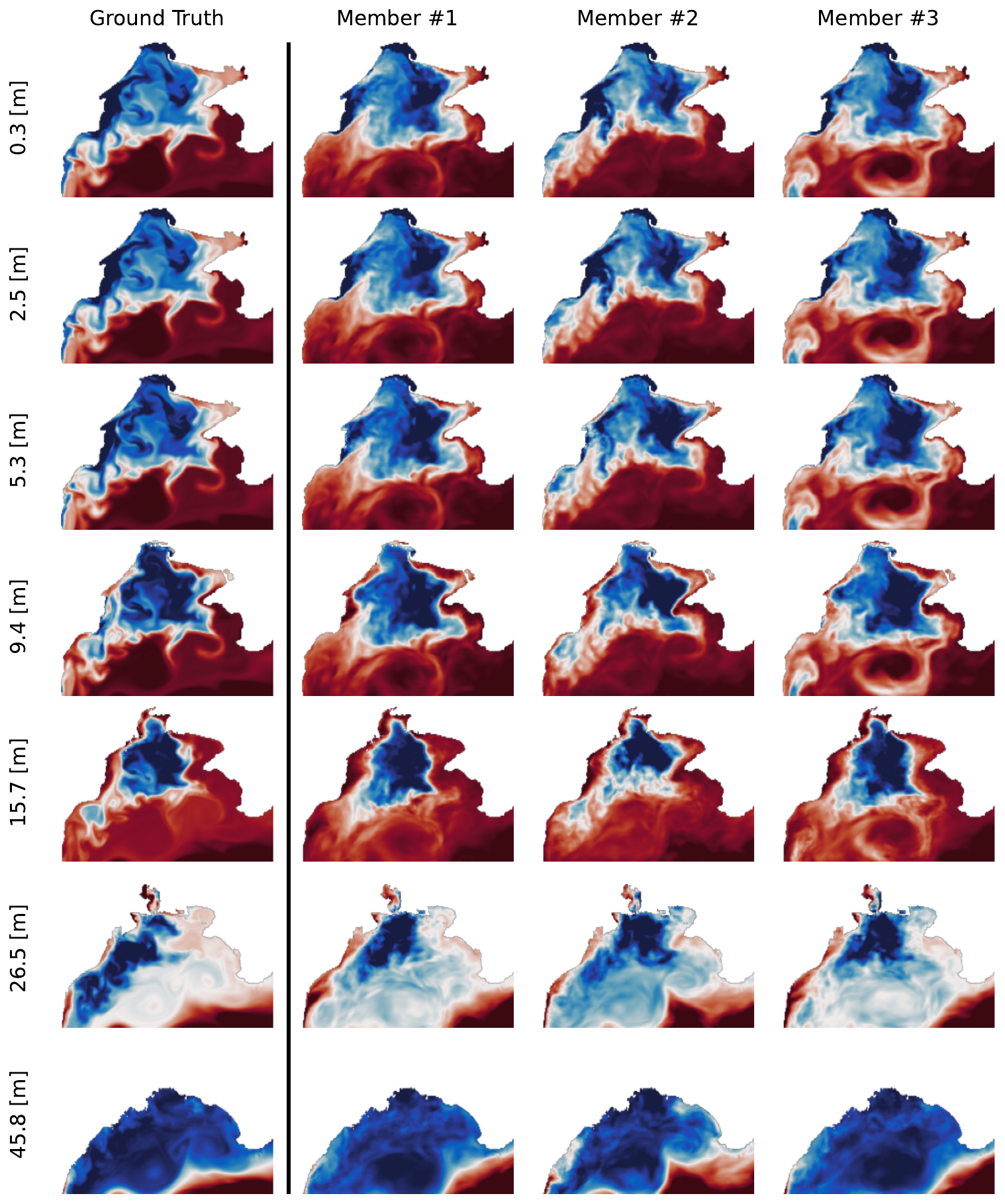}
    \caption{Posterior ensemble members $x \sim p_{\theta}(x \mid y, d)$ for \textbf{salinity} at various depths, using idealized observations $y$ shown in Fig.~\ref{fig-main-observations}A. The first column shows the ground truth state. The remaining columns show different ensemble members.}
    \label{fig-app-posterior-salinity}
\end{figure}

\newpage \begin{figure}[!h]
    \noindent\includegraphics[width=0.99\textwidth]{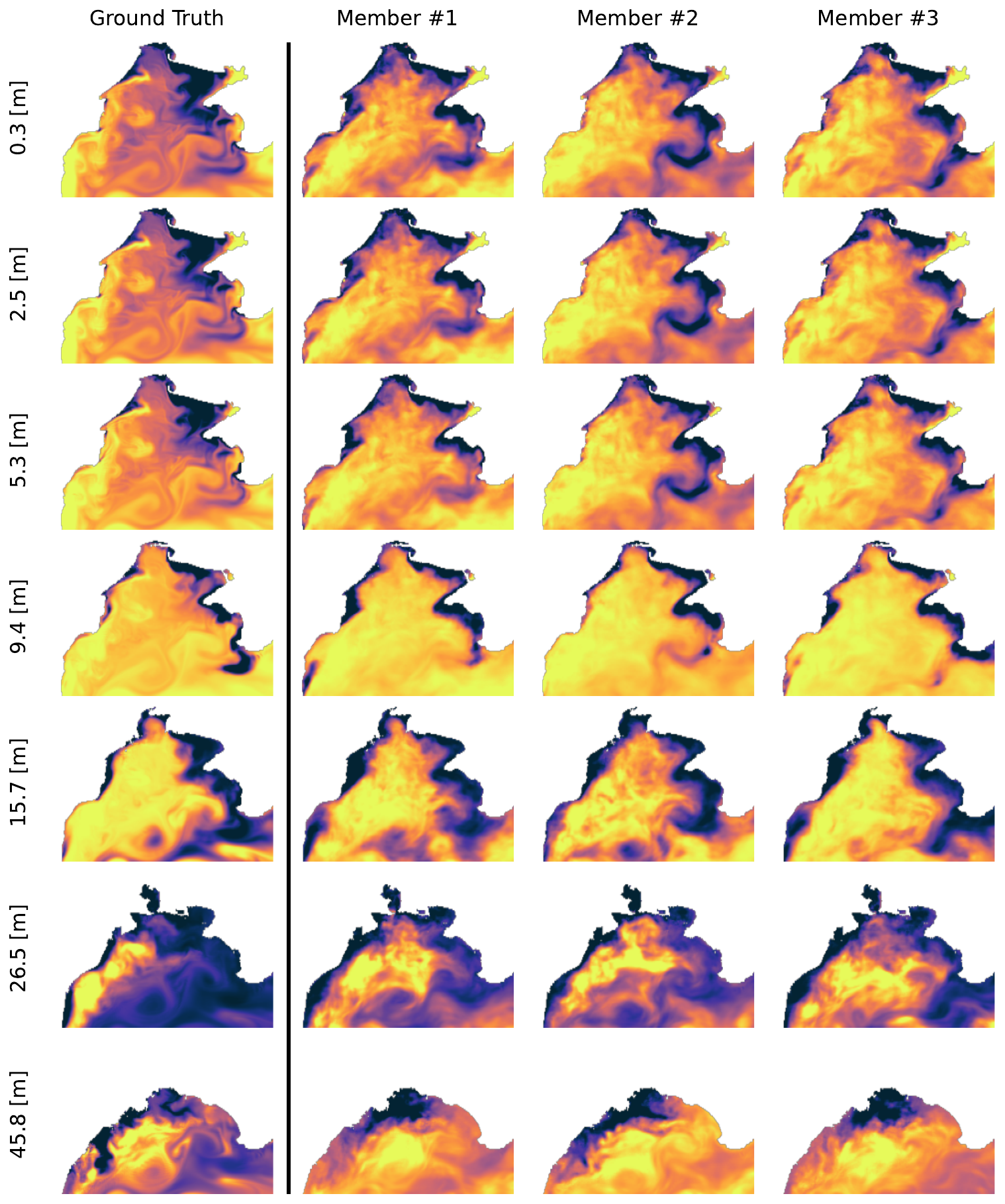}
    \caption{Posterior ensemble members $x \sim p_{\theta}(x \mid y, d)$ for \textbf{temperature} at various depths, using idealized observations $y$ shown in Fig.~\ref{fig-main-observations}A. The first column shows the ground truth state. The remaining columns show different ensemble members.}
    \label{fig-app-posterior-temperature}
\end{figure}

\newpage
\section{High-Resolution Observations}
\label{app-hr-observations}
\noindent We generate high-resolution observations $y^*$ with the observation model of Eq.~\eqref{eq-main-observation-process-approximation}, keeping the bias $\mu_{S}$ and the error covariance $\Sigma_{S}$ of Table~\ref{tab-app-observation-parameters}. The only difference lies in the observation operator: $A$ reduces to the identity for every variable, so each field is observed on the full simulation grid rather than on the coarser satellite grids. Such observations are not attainable with current satellites. They serve to quantify how much of the detection gap can be closed by improving the quality of surface observations alone.

\begin{figure}[!h]
    \noindent\includegraphics[width=0.99\textwidth]{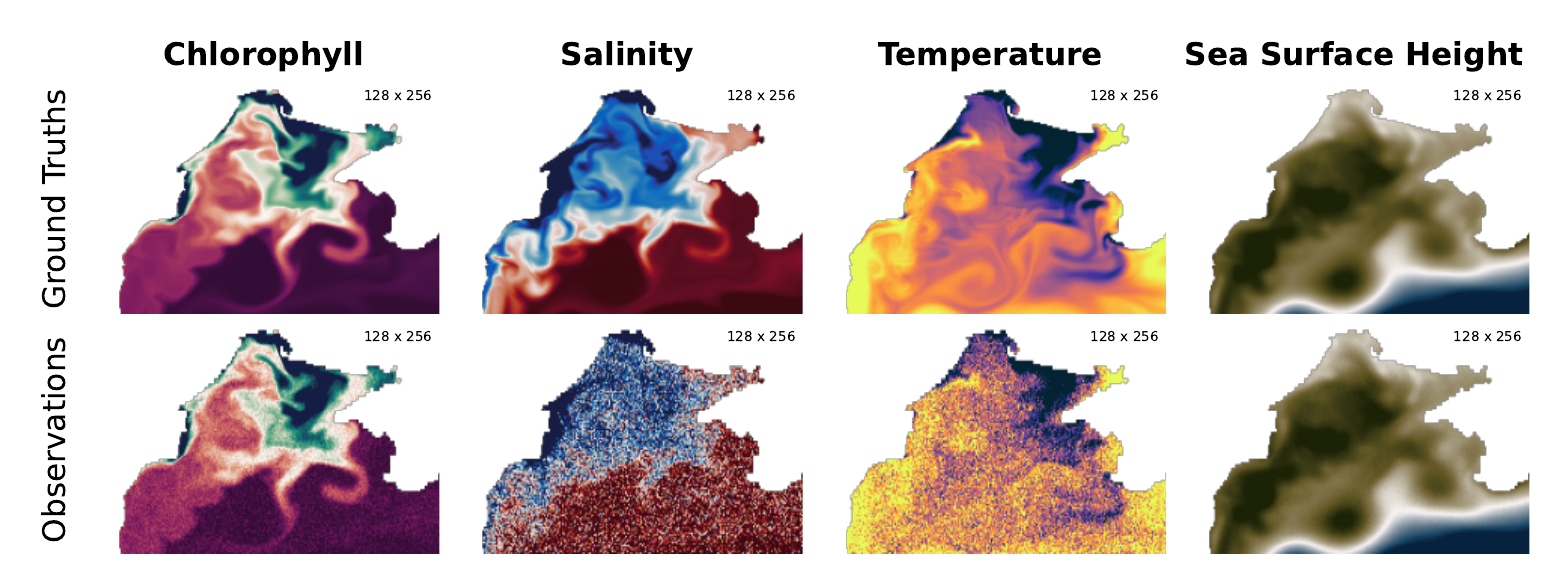}
    \caption{High-resolution Black Sea surface observations $y^*$ for the same state $x$ as in Fig.~\ref{fig-main-observations}A. All state variables are observed except oxygen.}
    \label{fig-app-hr-observations}
\end{figure}

\end{document}